\documentclass[12pt,letter]{article}
\usepackage[letterpaper, margin=1.3in]{geometry}

\usepackage{subfigure}
\usepackage{graphicx}
\usepackage{xcolor}
\usepackage{dcolumn}
\usepackage{color}
\usepackage{verbatim}
\usepackage{bm}       
\usepackage{graphicx}
\usepackage{epstopdf}
\usepackage[english]{babel}
\usepackage{amsmath}
\usepackage{amssymb}
\usepackage{booktabs}
\usepackage{multirow}
\usepackage{hhline}
\usepackage{tablefootnote}
\usepackage{threeparttable}

\usepackage{float}

\begin{document}

\title{On the dynamical stability of ferromagnetic \\ Ru and Os in the bct structure:\\ a first-principles study}

\author{M.E. Cifuentes-Quintal\thanks{Corresponding author. Email: miguel.cifuentes@cinvestav.mx  cifuentes.quintal@gmail.com}  and R. de Coss\\
\small\small Departamento de F\'isica Aplicada, \\
\small\small Centro de Investigaci\'on y de Estudios Avanzados del IPN, \\
\small\small Apartado Postal 73 Cordemex, 97310 M\'erida, Yucat\'an, M\'exico\\}

\date{\small\small }

\maketitle

\begin{abstract}

  Recent theoretical studies have predicted magnetic states for Ru and Os in the body centered tetragonal structure (bct) 
  with $c/a<1$. 
  In this study, we present first principles calculations of the phonon dispersion for ferromagnetic Ru- and Os-bct along the 
  epitaxial and uniaxial Bain paths, to evaluate their dynamical stability.
  The phonon dispersions were computed using the density functional perturbation theory, including the gradient corrections to
  the exchange-correlation functional within the plane-waves ultrasoft-pseudopotential approximation.
  The phonon dispersion for the local minimum in the Bain path with $c/a<1$ as well as the  uniaxial and epitaxial
  strained structures are analyzed. We find imaginary frequencies  along different directions of the Brillouin zone,
  which indicates that both systems are dynamically unstable.
  Consequently, ferromagnetic Ru and Os in the body centered tetragonal structure with $c/a<1$ are not truly metastable phases.

\end{abstract}

\section{Introduction}
The search for metals in metastable phases is a common approach to finding new materials. 
A metastable phase corresponds to a state in local mechanical equilibrium with a crystalline 
structure other than its ground-state, and it is therefore energetically less favorable \cite{metal-inst}.
Thus, one of the most active research fields in computational material science is the search for metastable structures 
with enhanced physical and chemical properties. 
Among these properties, ferromagnetism is one of the most important properties of matter because of its technological 
relevance and the low number of ferromagnetic elemental metals. 
Because the crystalline structure plays a major role in the definition of the properties in any material, 
a possible way to induce ferromagnetism in metals with paramagnetic ground-state is by changing its
crystalline structure to reach a metastable state.

Ruthenium (Ru) and Osmium (Os) are transition metals with hexagonal close packed structure (hcp) and paramagnetic ground-state.
Nevertheless, Kobayashi \textit{et al.} \cite{kobayashi} studied bulk Ru 
in the body-centered cubic structure (bcc) by means of first principles calculations using the APW method and the LDA approximation
for the exchange-correlation functional, and they find ferromagnetism for 5\% of the lattice constant 
expansion. Motivated by this result, Siiki and Hio \cite{shiiki} experimentally studied the formation of Ru thin-films  over 
substrates with different lattice constants, in an attempt to obtain ferromagnetic bcc Ru.  
Instead of obtaining the Ru-bcc structure, they find that the X-ray and electron beam diffraction patterns for Ru 
thin-films over Mo(110) suggest the formation of a body centered tetragonal structure (bct) with  lattice constants
of $a=3.24  \ {\rm \AA}$  and  $c=2.69  \ {\rm \AA}$ ($c/a=0.83$).
Subsequently, based on  first principles calculations using the FLAPW method and the GGA approximation
for the exchange-correlation functional, Watanabe \textit{et al.} \cite{watanabe} studied bulk Ru in 
the bcc, hcp, and face-centered cubic (fcc) structures as well as its tetragonal deformation path, which is better known as 
the uniaxial or classical Bain path. They determined that the total energy in the Bain path revealed a local minimum in the bct structure
for $c/a=0.83$ and a maximum in the bcc structure, which indicated that the Ru-bcc structure is unstable.
Recently, Sch\"{o}necker \textit{et al.} \cite{schonecker} using the FPLO 
method and the GGA approximation for the exchange-correlation functional, performed density functional theory (DFT) calculations  
of  the epitaxial Bain path for 28 non-magnetic ground-state transition metals to search for bulk ferromagnetic solutions
assuming that thick films could be coherently grown on substrates with fourfold symmetry.  
They find that Ru and Os in the bct structure with $c/a<1$ presents a ferromagnetic order.
Furthermore, Odkhuu \textit{et al.} \cite{odkhuu} employing  DFT calculations with the FLAPW method and the GGA approximation 
for the exchange-correlation functional, predicted  a large perpendicular magnetocrystalline anisotropy for bulk Ru in the bct structure,
which is two orders of magnitude greater than conventional magnetic metals.
These reports show the growing interest in the magnetic properties of Ru and Os in the bct structure.

It is important to note that in bulk metals, a crystalline  structure could correspond to a 
local minimum in the total energy along the Bain path, even though it could be dynamically unstable in that minimum \cite{metal-inst}. 
Moreover, based on DFT calculations it has been reported that cubic monoatomic metals are elastically unstable in the bct minimum of the Bain path 
 \cite{mehl}, which suggests that this result could also be the case for the hexagonal close packed metals.
Regardless of the prediction of a ferromagnetic order in the local minimum of the Bain path for Ru and Os in the bct
structure with $c/a<1$, 
 a study on the dynamical stability in these materials has not yet been reported to the best of our knowledge.
From a theoretical point of view,
 the stability analysis of a given crystal structure is necessary to evaluate if under small deformations of the lattice, 
the atoms return to their assumed equilibrium position, which identify a truly metastable phase \cite{metal-inst,mehl,aguayo,murrieta}. 

The aim of this study is to evaluate the dynamical stability of bulk ferromagnetic Ru and Os in the bct structure with 
$c/a<1$, using DFT-based first principles calculations. Thus, we determine the structural 
properties, magnetic moments, and phonon dispersion of Ru and Os 
for the local minimum in the Bain path for $c/a<1$ and for uniaxial and epitaxial strain in the range of $\pm$5 \%.
Imaginary phonon frequencies  were found along several directions of the Brillouin zone for all the
studied structures. Thus, the results indicate that bulk ferromagnetic Ru and Os in the bct structure are dynamically unstable.

\section{Computational details}
The DFT calculations were performed within the framework of the plane-waves pseudopotential  approximation (PW-PP), as implemented 
in the Quantum ESPRESSO package \cite{QE-2009}.
The exchange-correlation functional was treated with the PBE parametrization of the generalized gradient 
approximation \cite{PBE}.
The core electrons were replaced by scalar relativistic ultrasoft-pseudopotentials from the GBRV database \cite{GBRV}, 
and the valence wave functions and charge density
were expanded in plane waves with a kinetic energy cut-off of 40 and 400 Ry, respectively.

The integration on the irreducible part of the Brillouin zone was performed on a uniform grid of 24$\times$24$\times$24 
$k$-points, with a small cold smearing of 0.01 Ry \cite{mv-smearing}.
The phonon frequencies were computed by means of linear response theory, using the density functional perturbation theory \cite{DFPT} on a grid 
of 4$\times$4$\times$4 $q$-points and subsequently interpolated to obtain the full phonon dispersion.

\section{Results and discussion}

The bct crystalline structure belongs to the space group I4/mmm, with an in-plane lattice parameter $a$ and an out-of-plane lattice parameter $c$.
For the lattice vectors and the first Brillouin zone with its high symmetry points,  see the appendix A of reference \cite{setyawan}.
The lattice parameters for Ru and Os in the bct structure were obtained from the local minimum in their corresponding Bain paths for $c/a<1$.
In the epitaxial (uniaxial) Bain path, for each $a(c/a)$ the corresponding  $c/a(a)$ is obtained by a direct minimization 
of the electronic-total energy, thus making the out-of-plane (in-plane) stress disappear \cite{alippi}. 
The lattice parameters $a$ and $c$ in the local minimum for $c/a<1$ will be called $a_{bct}$ and $c_{bct}$, respectively.

\begin{table}[H]
\centering
\caption{Lattice parameters $a_{bct}$ and $(c/a)_{bct}$, energy of the bct structure with respect to the hcp ground-state ($E_{bct}-E_{hcp}$), 
and magnetic moments (MM) for Ru and Os in the bct structure. Previously reported values are included for comparison.}
  \begin{threeparttable}
\begin{tabular}{ccccccc}
\hline
\hline

\multirow{2}{*}{\textrm{Metal}} & 
\multicolumn{1}{c}{$a_{bct}$}  & 
\multirow{2}{*}{$(c/a)_{bct}$}   & 
\multicolumn{1}{c}{\textrm{$E_{bct}-E_{hcp}$}} & 
\multicolumn{1}{c}{\textrm{MM}} &
\multirow{2}{*}{\textrm{Method}}  &
\multirow{2}{*}{\textrm{Ref.}}  \\

\textrm{}  & 
\multicolumn{1}{c}{\textrm{(\AA)}} &                
\textrm{}  & 
\multicolumn{1}{c}{\textrm{(mRy/atom)}} & 
\multicolumn{1}{c}{\textrm{($\mu_B$/atom)}} &
\textrm{}  &
\textrm{}  \\

\hline
\multirow{5}{*}{\textrm{Ru}}    &   3.24   &   0.83   &  -  &  -   & Exp.\tnote{a} &  \cite{shiiki} \\
      &   3.25   &   0.83   &  40   &  0.40  & FLAPW &  \cite{watanabe}\\
      &   3.18   &   0.85   &  44  &   0.62  & FPLO &  \cite{schonecker}  \\
      &   3.25   &   0.84   &  35   &  0.60  &   FLAPW &\cite{odkhuu} \\
      &   3.24   &   0.84   &  38   &  0.62  &  PW-PP & This work  \\
\hline
\multirow{2}{*}{\textrm{Os}}    &   3.30   &   0.84   &  53   &  0.50   & FPLO &  \cite{schonecker} \\
      &   3.30   &   0.83   &  56   &  0.47    & PW-PP & This work \\

\hline		
\hline
\end{tabular}
\begin{tablenotes}
\item[a] {\small Data for Ru thin-film prepared by ion beam sputtering on Mo(110) single-crystal wafer.}
     \end{tablenotes}
  \end{threeparttable}
\label{par-estr}
\end{table}

The calculated structural properties and the magnetic moments of Ru and Os in the bct structure are summarized in Table \ref{par-estr}.
For the lattice parameters, we find a good agreement with previously reported 
values  \cite{shiiki,watanabe,schonecker,odkhuu}. 
Considering the calculated energy of the bct structure with respect to the hcp ground-state energy ($E_{hcp}$),
we determine that Ru and Os have the highest energy difference $E_{bct}-E_{hcp}$, among the transition metals with 
an hcp ground-state  \cite{bct-hcp}.
The calculated value of the magnetic moment for Ru in the bct structure is $0.62 \ \mu_B/{\rm atom}$, which is in good agreement with recent 
reports  \cite{schonecker,odkhuu}; however, this value is larger than the computed value by Watanabe \textit{et al.} \cite{watanabe} of  $0.40 \ \mu_B/{\rm atom}$.
The obtained magnetic moment for Os (0.47 \ $\mu_B/{\rm atom}$) is in good agreement with the value of 0.50 \ $\mu_B/{\rm atom}$ reported by
Sch\"{o}necker \textit{et al.} \cite{schonecker}.

\begin{figure}
\centering
 \includegraphics*[scale=1]{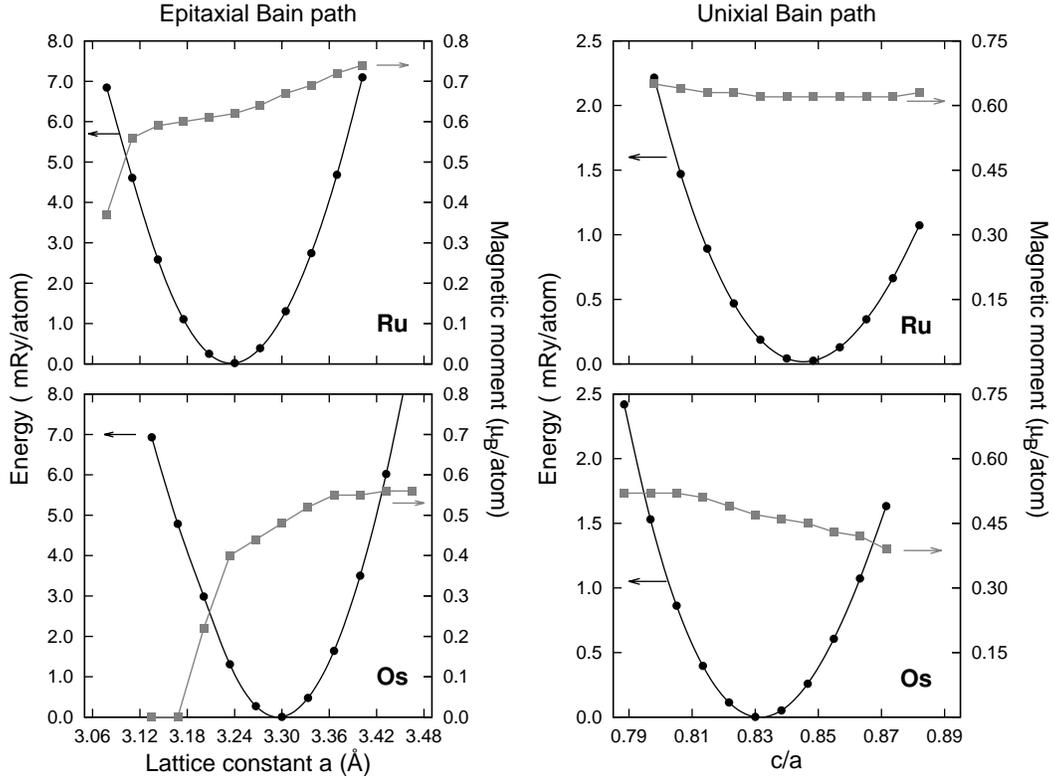}
 \caption{Epitaxial and uniaxial Bain paths for Ru and Os in the bct structure around the local minimum for $c/a<1$. 
 The energy scale is respect to $E_{bct}$.}
\label{all-paths}
\end{figure}

In Fig. \ref{all-paths} we present the epitaxial and uniaxial Bain paths for Ru and Os
around the local minimum for $c/a<1$, including the evolution of the magnetic moment for each path.
As it can be observed, the epitaxial deformation produces energy and magnetic moment changes that are larger than the uniaxial 
deformation. 
This behavior can be explained considering that the volume per atom is given by $V=a^2 c/2$.
Thus, for a given material in the bct structure, the deformations on the $a$ lattice parameter produce a larger effect on $V$ than  deformations in the $c$ 
lattice parameter, and consequently on the total energy and the magnetic moment \cite{kubler}.
It is interesting to note that the magnetic moment for Os in the bct structure is more sensitive to the epitaxial and uniaxial strains
than for Ru in the bct structure.
In fact, we find that the magnetic moment in Os changes from 0 to nearly $0.6 \ \mu_B/{\rm atom}$ in the epitaxial strain range of 
$-4$ to $+4\%$. 
This type of behavior is quite interesting because of the possibility of developing a strain engineering of the 
magnetic moment in the Os-bct structure. 

\begin{figure}
\centering
 \includegraphics*[scale=1]{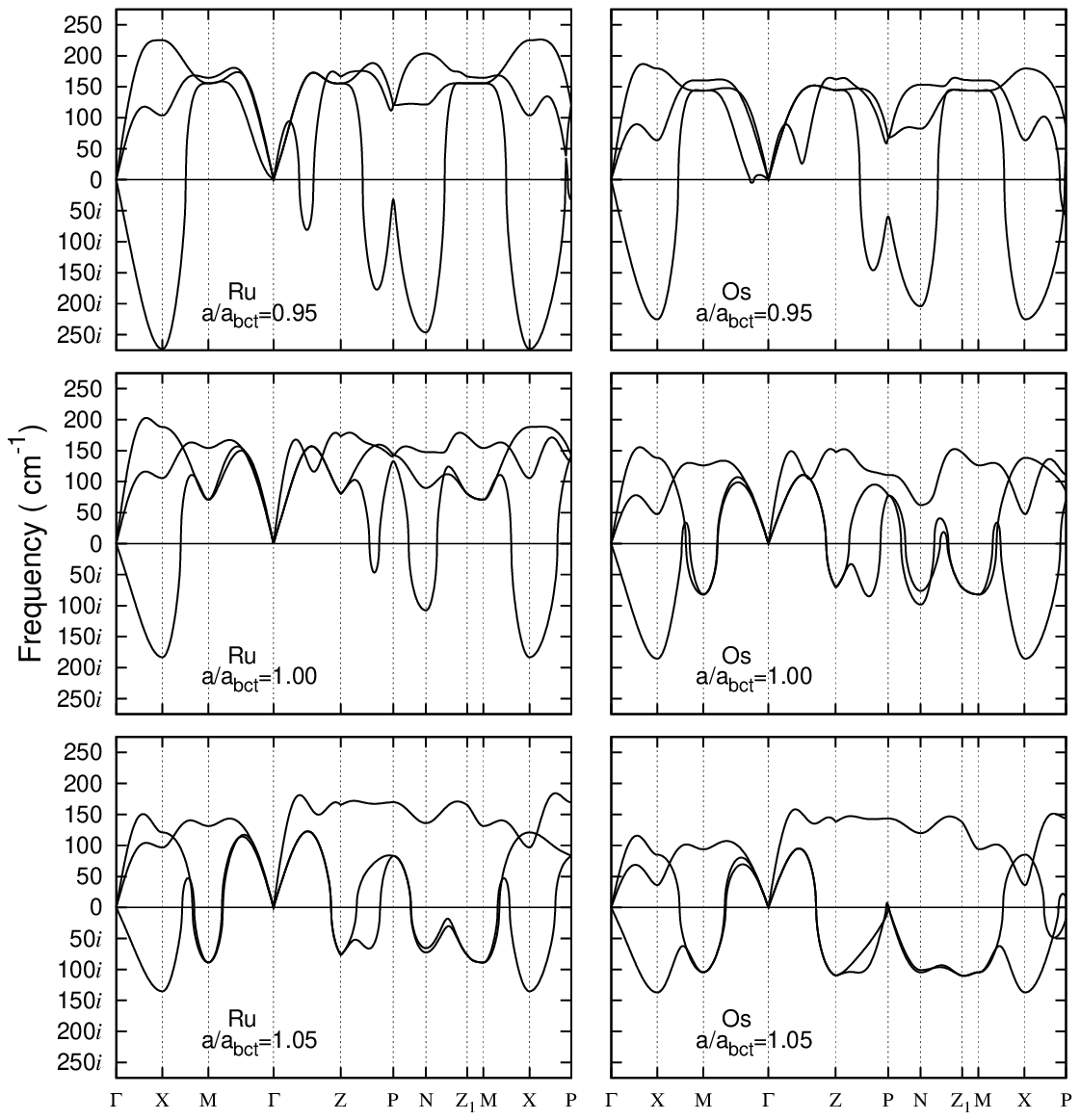}
 \caption{Phonon dispersion for Ru and Os in the bct structure along the epitaxial Bain path plotted in Fig. \ref{all-paths}.}
\label{phdisp-epitax}
\end{figure}
\begin{figure}
\centering
 \includegraphics*[scale=1]{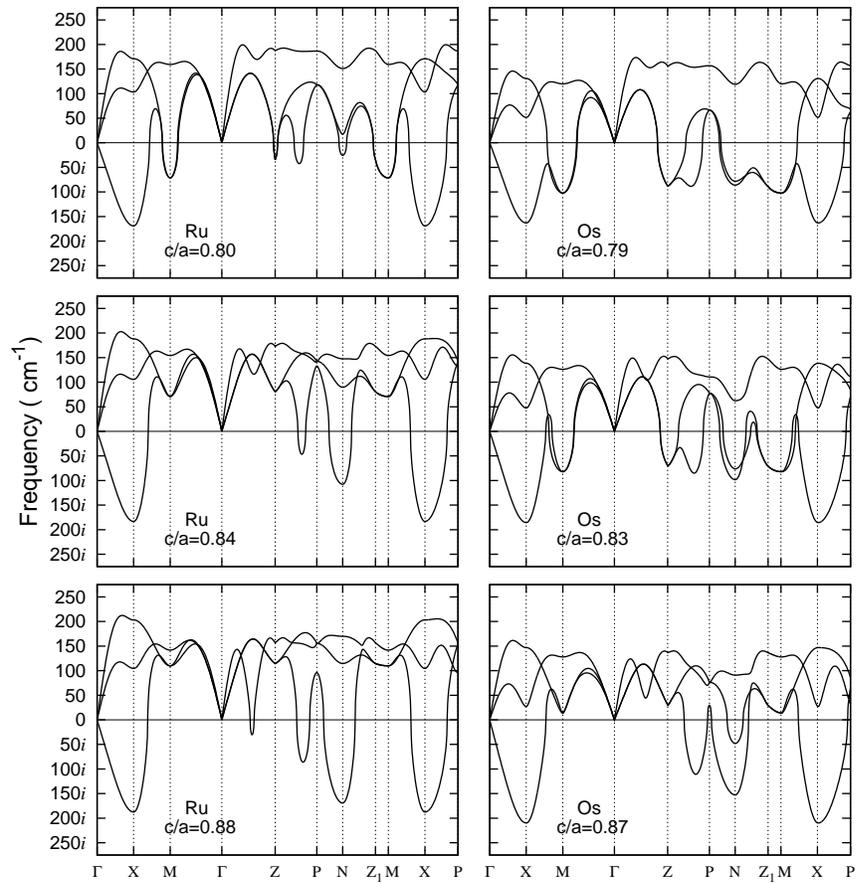}
\caption{Phonon dispersion for Ru and Os in the bct structure along the uniaxial Bain path plotted in Fig. \ref{all-paths}.}
\label{phdisp-uniax}
\end{figure}

To evaluate the dynamical stability of the ferromagnetic Ru- and Os-bct structures around the local minimum for $c/a<1$, 
we have calculated the phonon dispersion in the epitaxial and uniaxial Bain paths.
In Figs. \ref{phdisp-epitax} and \ref{phdisp-uniax}, we present the phonon dispersion for Ru and Os in the bct structure 
at the local minimum and $\pm 5\%$ of the epitaxial and uniaxial strain, respectively. 
In all the cases, we find imaginary phonon frequencies that clearly indicate the dynamic instability of bulk Ru and Os, in the bct 
structure. It should be noted that the imaginary phonon  branch in the $\Gamma \rightarrow {\rm X}$ direction reveals an
elastic instability related to a negative value of the elastic constant $C'=(c_{11}-c_{12})/2$  \cite{metal-inst,mehl,marcus}. 
Additionally, the effect of the epitaxial and uniaxial strains on the phonon dispersion was evaluated.  
It was determined that neither the epitaxial strain nor the uniaxial strain contribute to the stability of the bct structure in
both metals, Ru and Os.
Furthermore, due to the high values of the imaginary phonon frequencies, it is unlikely that the temperature 
effects can reverse that instability. 
Consequently, ferromagnetic Ru and Os in the body centered tetragonal structure with $c/a<1$ are not truly metastable phases.

In order to analyze the contribution of the magnetic ordering to the dynamical instability of Ru and Os in the body centered tetragonal 
structure, in addition to the ferromagnetic order (FM), we also consider antiferromagnetic order (AF) and the non-magnetic system (NM). 
The total energy calculations for Ru and Os in the NM and AF state shows a local minimum along the Bain path for  $c/a<1$ (not shown here),
with lattice parameters very close to the corresponding FM state. Nevertheless, for both metals the energy of the local minimum in the NM and AF states are 
higher than the FM state. For Ru(Os), the AF-FM and NM-FM energy differences are 2.1(0.7) and 2.4(0.8) mRy/atom, respectively.

\begin{figure}
\centering
 \includegraphics*[scale=1]{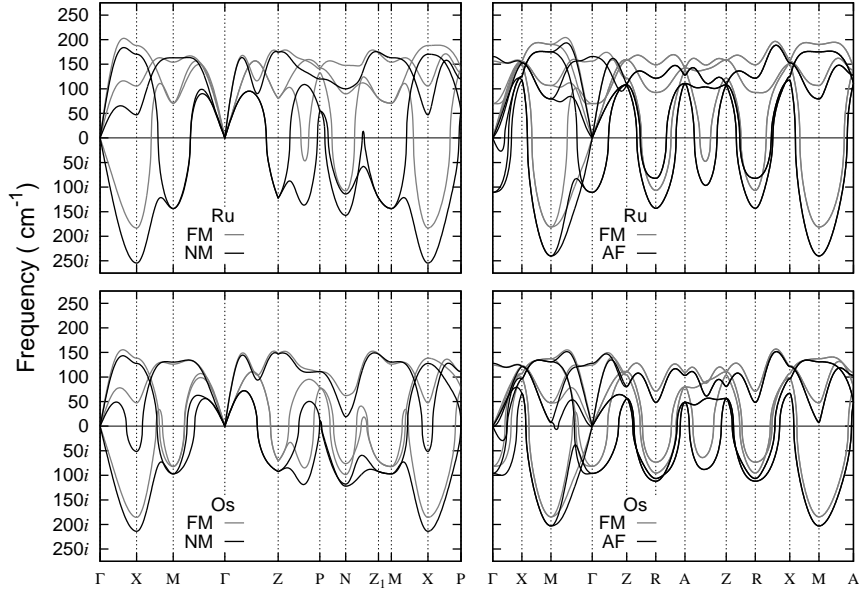}
\caption{Phonon dispersion for Ru and Os in the bct structure for the NM, FM, and AF order. The
 AF calculations were performed using the simple tetragonal unit cell with a two atom basis.}
\label{phdisp-magn}
\end{figure}

Considering that the structural parameters for the three studied magnetic states (NM, AF and FM) are very similar, 
we have calculated the phonon dispersions for the NM and AF states at the structural lattice parameters of the FM state. 
In this way we are considering only the effect of the magnetic ordering on the phonon instabilities.
In Fig. \ref{phdisp-magn} we show the phonon dispersion for Ru and Os in the body centered tetragonal structure with the NM and AF order, 
including the previous FM state for reference. Thus, we found  that the NM and AF states are dynamically instable for both metals.  
Furthermore, the values of the imaginary phonon frequencies for the NM and AF states are larger than the FM state.  These results show 
that the ferromagnetism is not the origin of the instability of Ru and Os in the body centered tetragonal structure.

Finally, we want to emphasize that the predicted dynamical instability of ferromagnetic Ru and Os in the bct 
structure applies to bulk.
Thus, the implication of this study is that there can be no free-standing bct structures with $c/a<1$
for these metals. Any bct structures with a $c/a<1$ observed for Ru and Os, such as the thin films of Ru on
Mo(110) \cite{shiiki}, must be stabilized by an external force. 
In the case of ultrathin films, the material is stabilized by external effects, such as surface energy, 
substrate strain, low dimensionality and finite size, among others.
From a theoretical point of view, a correct first principles prediction on the stability and the properties of
transition metals in ultrathin films should explicitly include the overlayers and the substrate, and cannot
be based only on bulk calculations, particularly when the bulk structure of the metal-overlayer is dynamically
unstable.

\section{Summary}

We have performed first principles calculations of the phonon dispersion for ferromagnetic
Ru and Os in the bct structure, to evaluate their  dynamical stability.
The phonon dispersion for the local minimum in the Bain path with $c/a<1$ and the uniaxial and epitaxial
strained structures were analyzed. 
For both systems, we find imaginary frequencies  along different directions of the Brillouin zone, which
indicated dynamical instability.
This result shows that ferromagnetic Ru and Os in the body centered tetragonal structure with $c/a<1$ are not 
truly metastable phases.

\section*{Acknowledgements}
One of the authors (M.E.C.-Q.) gratefully acknowledges a student grant from CONACYT-M\'exico.
Computational resources were provided by ``Cluster H\'ibrido de Superc\'omputo - Xiuhcoatl" at
Cinvestav. This study was partially supported by CONACYT-M\'exico. 
Finally, the authors gratefully acknowledges to the anonymous referees for constructive comments.


\begin{thebibliography}{10}
\expandafter\ifx\csname url\endcsname\relax
  \def\url#1{\texttt{#1}}\fi
\expandafter\ifx\csname urlprefix\endcsname\relax\def\urlprefix{URL }\fi
\expandafter\ifx\csname href\endcsname\relax
  \def\href#1#2{#2} \def\path#1{#1}\fi

\bibitem{metal-inst}
G.~Grimvall, B.~Magyari-K\"ope, V.~Ozoli\ifmmode \mbox{\c{n}}\else
  \c{n}\fi{}\ifmmode~\check{s}\else \v{s}\fi{}, K.~A. Persson, Rev. Mod. Phys. 84 (2012)
  945.

\bibitem{kobayashi}
M.~Kobayashi, T.~Kai, N.~Takano, K.~Shiiki, J. Phys.: Condens. Matter 7 (1995) 1835.

\bibitem{shiiki}
K.~Shiiki, O.~Hio, Jpn.  J. Appl. Phys. 36 (1997) 7360.

\bibitem{watanabe}
S.~Watanabe, T.~Komine, T.~Kai, K.~Shiiki, J. Magn. Magn. Mater. 220 (2000) 277.


\bibitem{schonecker}
S.~Sch\"onecker, M.~Richter, K.~Koepernik, H.~Eschrig,  Phys. Rev. B 85 (2012) 024407.
  
 
\bibitem{odkhuu}
D.~Odkhuu, S.~Rhim, N.~Park, K.~Nakamura, S.-C. Hong, Phys. Rev. B 91 (2015) 014437.

\bibitem{mehl}
M.~J. Mehl, A.~Aguayo, L.~L. Boyer, R.~de~Coss, Phys. Rev. B 70  (2004) 014105.

\bibitem{aguayo}
A.~Aguayo, G.~Murrieta, R.~de~Coss, Phys. Rev. B 65  (2002) 092106.

\bibitem{murrieta}
G.~Murrieta, A.~Tapia, R.~de~Coss, Carbon 42  (2004) 771.

\bibitem{QE-2009}
P.~Giannozzi, et~al.,   J. Phys.: Condens. Matter 21 (2009)  395502.

\bibitem{PBE}
J.~P. Perdew, K.~Burke, M.~Ernzerhof, Phys. Rev. Lett. 77 (1996) 3865.

\bibitem{GBRV}
K.~F. Garrity, J.~W. Bennett, K.~M. Rabe, D. Vanderbilt, Comp. Mater. Sci. 81 (2014) 446.

\bibitem{mv-smearing}
N.~Marzari, D.~Vanderbilt, A.~De~Vita, M.~C. Payne, Phys. Rev. Lett. 82 (1999) 3296.

\bibitem{DFPT}
S.~Baroni, S.~de~Gironcoli, A.~Dal~Corso, P.~Giannozzi, Rev. Mod. Phys. 73 (2001) 515.

\bibitem{setyawan}
W.~Setyawan, S.~Curtarolo, Comp. Mater. Sci. 49 (2010) 299.

\bibitem{alippi}
P.~Alippi, P.~Marcus, M.~Scheffler, Phys. Rev. Lett. 78 (1997) 3892.

\bibitem{bct-hcp}
Our calculated values of $E_{bct}-E_{hcp}$ for Sc, Y, Ti, Zr, Hf, Tc, and Re:
8.2, 8.9, 4.5, 11.1, 17.6, and 22.6 mRy/atom, respectively.


\bibitem{marcus}
P.~M. Marcus, F.~Jona, S.~L. Qiu, Phys. Rev. B 66 (2002) 064111.

\bibitem{kubler}
J.~K\"ubler, \textit{Theory of Itinerant Electron Magnetism}, Vol. 106, Oxford University Press, New York, 2000, pp. 194-198.
  
\end{thebibliography}
\end{document}